\documentclass[aps,prd,floatfix,showpacs,10pt,nofootinbib]{revtex4-1}

\usepackage{amsmath,amssymb,amsfonts,amsthm}
\usepackage{color}
\usepackage{float}
\usepackage{graphicx}
\usepackage{pstricks,pstricks-add}
\usepackage{pst-plot}
\usepackage[scriptsize,nooneline,hang]{caption}
\usepackage[hang,nooneline,scriptsize]{subfigure}

\definecolor{dark-green}{rgb}{0,0.7,0}
\definecolor{dark-blue}{rgb}{0,0.2,0.5}
\definecolor{med-blue}{rgb}{0,0.7,1}
\definecolor{mblue}{rgb}{0,0.2,1}
\definecolor{cnc}{rgb}{0.8,0,0}
\definecolor{light-red}{rgb}{1,0.8,0.8}
\definecolor{dark-yellow}{rgb}{1,0.8,0}
\definecolor{light-blue}{rgb}{0.8,0.9,1}
\definecolor{grey}{rgb}{0.211,0.211,0.211}
\definecolor{verylight-blue}{rgb}{0.93,0.95,1}
\definecolor{light-yellow}{rgb}{1,0.9,0.8}

\newcommand{\weglassen}[1]{}

\begin{document}

\title{ Gravitating non-Abelian cosmic strings}

\author{Ant\^ onio de P\'adua Santos}
\email{padua.santos@gmail.com}

\author{Eug\^enio R. Bezerra de Mello  }
\email{emello@fisica.ufpb.br}

\affiliation{
Departamento de F\'{\i}sica, Universidade Federal da Para\'{\i}ba, 58.059-970, 
Caixa Postal 5.008, Jo\~{a}o Pessoa, PB, Brazil}

\date\today

\begin{abstract}
In this paper we study regular cosmic string solutions of the non-Abelian Higgs model coupled with  
gravity. In order to develop this analysis, we constructed a set of coupled non-linear differential equations.
Because there is no closed solution for this set of equations, we solve it 
numerically. The solutions that we are interested in asymptote to a flat space-time 
with a planar angle deficit. This model under consideration present two bosonic sectors,
besides the  non-Abelian gauge field. The two bosonic
sectors may present a direct coupling. So, we investigate the relevance of this coupling on
the system, specifically in the linear energy density of the string and on the 
planar angle deficit. We also analyze the behaviors of these quantities
as function of the energy scale where the gauge symmetry is spontaneously broken.
\end{abstract}


\maketitle

\section{Introduction}
\label{Int}

According to the Big Bang theory, the universe has been expanding and cooling. During its early evolution, 
the universe has been underwent a series of phase transitions characterized by spontaneous symmetry 
breaking \cite{Kibble1976}. These phase transitions are important, because they 
provide a mechanism for the formation of topological defects such as domain 
walls, monopoles, cosmic string, among others  \cite{Vilenkin-Shellard}.
In the eighties of last century  the interest to cosmic strings has been increase due to
the fact  that they were considered  candidates to provide a mechanism for the
large-scale structure formation in the universe. Although the recent observations of the
Cosmic Microwave Background (CMB) radiation have ruled out cosmic strings as 
seed for the primordial density perturbations, they are still candidates to explain 
a small non-Gaussianity in the cosmic microwave background and the influence on the temperature anisotropies. 
Such effects had origins in the gravitational fields of cosmic strings in motion \cite{Ade:2013}.
In addition, astrophysical and cosmological consequences of formation of 
strings are emission of gravitational waves and high energy cosmic
rays such as neutrinos and gamma-rays. These observational data can 
help to constraint the product of the Newton's constant $G$ to
the linear mass density of cosmic string, $\mu$ \cite{Hindmarsh}. These effects 
are generated by cosmic strings formed under inflation 
context\footnote{Observational consequences generated by 
cosmic strings can be found in \cite{Hindmarsh, Copeland}.} which 
makes the physics of cosmic strings a vast area of interest.

The first theoretical model associated with strings-like solutions  was given  
by Nielsen and Olesen \cite{N-O} by   using an Abelian Higgs model  lagrangian
which presents a spontaneously $U(1)$ gauge symmetry broken.  These solutions 
were also named by {\bf vortex}. The influence of this system on the geometry of the spacetime
has been analyzed in \cite{DG, Laguna} many years ago, by coupling the energy-momentum tensor associated with
the Nielsen and Olesen string with the Einstein equations.  In these papers, static and cylindrically 
solutions for the matter and  gauge  fields and also for the metric tensor,  were investigated numerically  by
using coupled ordinary non-linear differential equations. A few years latter the Abelian Higgs model has also investigated 
 in \cite{Christensen} and \cite{Brihaye-Lubo}. The so-called (p,q) gravitating cosmic string have been
 analyzed numerically in \cite{Betti}. In the latter, static and cylindrically solutions  of the system 
 containing two different Abelian gauge fields coupled with two bosonic sectors were investigated.

In this paper we are interested to analyze the cosmic string solution again, however, considering 
at this time the $SU(2)$  non-Abelian Higgs model coupled to gravity.
As  it was also shown in \cite{N-O}, to create a non-Abelian topological stable string, it is necessary the presence of two bosonic
iso-vectors coupled with the gauge field. The  interaction potential should present  terms with fourth powers in
these bosonic fields in order to have spontaneously gauge symmetry broken. Moreover, it 
may contain a direct interaction between these bosonic sectors.  
 
This paper is organized as follows: In section \ref{Model} we  present the non-Abelian Higgs model coupled to gravity
and analyze the conditions  that the physical parameters contained in the potential should satisfy in order the 
system provides stable topological solutions. In section \ref{Ansatz} we 
present the ansatz for the matter and gauge fields, and for the  metric tensor. The equations of motion and boundary 
conditions obeyed by the fields and metric tensor are presented in section \ref{Equation}. In section \ref{Numerical} we exhibit 
our numerical results for the behaviors of the fields considering different values of the relevant parameters. 
In this section we present a comparison of this non-Abelian system with the Abelian one.
Also we investigate the behaviors of the energy density of the string and planar angle deficit, as functions
of the energy scale where the $SU(2)$ gauge symmetry is spontaneously broken and the
interaction  coupling  between the bosonic sectors.
In section \ref{Concl} we give our conclusions and in the Appendix we present a special 
solution of the system that, although the corresponding configuration is not in according to stable topological
configuration, it behaves as a BPS solution of the fields equations .

\section{The model}
\label{Model}

The model that we want to study is described by the following action, $S$:
\begin{equation}
 S = \int d^4x \sqrt{-g}\left(\frac{1}{16\pi G}R + \mathcal{L}_m\right), \label{eqAction}
\end{equation}
 where $R$ is the Ricci scalar, $G$ denotes the Newton's constant and $ \mathcal{L}_m $ is the matter Lagrangian density of the 
non-Abelian Higgs model given by
\begin{equation}
 \mathcal{L}_m = -\frac{1}{4}F^a_{\mu \nu}F^{\mu \nu a} + \frac{1}{2}(D_{\mu}\varphi^a)^2 + 
 \frac{1}{2}(D_{\mu}\chi^a)^2 - V(\varphi^a, \chi^a),
 \quad a = 1, 2, 3. \label{eqLagrangian}
\end{equation}
The field strength tensor is 
\begin{equation}
 F^{a}_{\mu \nu} = \partial_{\mu} A^a_{\nu} - \partial_{\nu} A^a_{\mu} + e \epsilon^{abc}A^b_{\mu}A^c_{\nu},
\end{equation}
with the $SU(2)$ gauge potential $A^b_{\mu}$, and $e$ being the gauge coupling constant. 
The covariant derivatives of the Higgs fields are given by  \cite{N-O},
\begin{equation}
 D_{\mu}\varphi^a = \partial_{\mu} \varphi^a + e \epsilon^{abc}A^b_{\mu}\varphi^c,
\end{equation}
\begin{equation}
 D_{\mu}\chi^a = \partial_{\mu} \chi^a + e \epsilon^{abc}A^b_{\mu}\chi^c,
\end{equation}
where the latin indices denote the internal gauge groups $(a,b = 1,2,3)$. The interaction 
potential, $V(\varphi^a, \chi^a)$,  which we shall consider is expressed by
\begin{eqnarray}
V(\varphi^a, \chi^a) &  =  &\frac{\lambda_1}{4}\left[(\varphi^a)^2 - \eta_1^2\right]^2 
+ \frac{\lambda_2}{4}\left[(\chi^a)^2- \eta_2^2\right]^2 \nonumber \\
& & + \frac{\lambda_3}{2}\left[(\varphi^a)^2 - \eta_1^2\right]\left[(\chi^a)^2 - \eta_2^2\right],
\end{eqnarray}
where the $\lambda_1 $ and $\lambda_2 $ are the Higgs fields self-coupling positive constants 
and $\lambda_3 $ is the coupling constant between both  bosonic
sectors. $\eta_1 $ and $\eta_2$ parameters correspond to energy scale where the gauge 
symmetry is broken. The potential above has different 
properties according to the sign of $\Delta \equiv \lambda_1 \lambda_2 - \lambda_3^2$:
\begin{itemize}
 \item For $\Delta>0 $, the potential has positive value and its minimum is attained 
 for $(\varphi^a)^2 = \eta_1^2 $ and $(\chi^a)^2 = \eta_2^2$.
  \item For $\Delta<0$, these configuration becomes saddle points and two minima occur for:
\begin{equation}
 (\varphi^a)^2 = 0, \quad (\chi^a)^2 = \eta_2^2 + \frac{\lambda_3}{\lambda_2}\eta_1^2
\end{equation}
and 
\begin{equation}
 (\chi^a)^2 = 0, \quad (\varphi^a)^2 = \eta_1^2 + \frac{\lambda_3}{\lambda_1}\eta_2^2.
\end{equation}
The values of the potential for these cases are, respectively, 
\begin{equation} 
V_{min}=\frac{\eta_1^4}{4\lambda_2}\Delta \quad \textrm{and} \quad V_{min}=\frac{\eta_2^4}{4\lambda_1}\Delta.
\end{equation}
Both values for $V_{min} $ are negatives, since $\Delta<0$.  
\end{itemize}

\subsection{The Ansatz}
\label{Ansatz}
The Euler-Lagrange equations for the Higgs  and
gauge fields,  and the Einstein equations for the metric tensor are obtained as presented below. 

First let us consider the most general, 
cylindrically symmetric line element invariant under boosts along z-direction. 
By using cylindrical coordinates, this line element is given by:
\begin{equation}
 ds^2 = N^2(\rho)dt^2 - d\rho^2 - L^2(\rho)d\phi^2 - N^2(\rho)dz^2 \  .
 \label{ds}
\end{equation}
For this metric, the only non-vanishing components of the Ricci tensor, $R_{\mu\nu}$,  are:
\begin{equation}
 R_{tt} = - R_{zz} = \frac{NLN''+ NN'L' + L(N')^2}{L} \  ,
\end{equation}
\begin{equation}
 R_{\rho\rho} = \frac{2LN'' + NL''}{NL} \  ,
\end{equation}
\begin{equation}
 R_{\phi \phi} = \frac{L(2N'L' + NL'')}{N} \  ,
\end{equation}
where the primes denotes derivative with respect to $\rho$.

As to the  bosonic sectors and gauge fields \cite{Vega}, we use the expressions below:
\begin{equation}
 \varphi^a(\rho) = f(\rho)
\left( \begin{array}{c}
 \cos(\phi)\\ 
 \sin(\phi)\\
   0
\end{array} \right) \  , 
\end{equation}
\begin{equation}
 \chi^a(\rho) = g(\rho)
\left( \begin{array}{c}
 -\sin(\phi)\\ 
 \cos(\phi)\\
   0
\end{array} \right)  \  ,
\end{equation}
\begin{equation}
\vec{A}^a(\rho) = \hat{\phi}\left(\frac{1-H(\rho)}{e\rho}\right)\delta_{a,3} =
 -\hat{\phi}\frac{A(\rho)}{e\rho}\delta_{a,3}  
\end{equation}
and 
\begin{equation}
 {A}^a_t(\rho) = 0 \  , \quad  a = 1, 2,3 \  .
\end{equation}
We can see that both iso-vector bosonic fields satisfy the orthogonality condition,  $\varphi^a\chi^a=0$.

\section{Equation of motion}
\label{Equation}

In order to present the equations of motion in a more appropriate form to apply numerical analyzes, we shall 
define new dimensionless variables and functions as shown below:
\begin{equation}
 x = \sqrt{\lambda_1}\eta_1\rho, \quad f(\rho) = \eta_1X(x), \quad g(\rho) = \eta_1Y(x) \quad {\rm and}
   \ \  L(x) = \sqrt{\lambda_1}L(\rho)\eta_1 \  .
\end{equation}
Therefore, the Lagrangian depends only on  dimensionless variables and parameters: 
\begin{equation}
 \alpha = \frac{e^2}{\lambda_1}, \quad q = \frac{\eta_1}{\eta_2}, \quad \beta^2_i = 
 \frac{\lambda_i}{\lambda_1}, \quad i = 1, 2, 3 \  ,  \
\gamma = \kappa \eta_1^2 \ {\rm and} \  \kappa = 8\pi G \ .
\end{equation}
It is convenient to use the Einstein equation in the form
\begin{equation}
 R_{\mu\nu} = -\kappa\left(T_{\mu\nu} -\frac{1}{2}g_{\mu\nu}T\right), 
 \quad \textrm{with} \quad T = g^{\mu\nu}T_{\mu\nu}  \quad 
\textrm{and}\quad \mu, \nu = t, x, \phi, z \  .  \label{eqEinstein}
\end{equation}
The energy-momentum tensor associated with the mater field is defined by,
\begin{equation}
T_{\mu\nu} = \frac{2}{\sqrt{-g}}\frac{\delta S}{\delta g^{\mu\nu}}, \quad g = \textrm{det}(g_{\mu\nu}).
\end{equation}

Varying the action  (\ref{eqAction}) with respect to matter fields and metric tensor, we obtain a system of 
five non-linear coupled differential equations.  The Euler-Lagrange equations are:
\begin{equation}
 \frac{(N^2LX')'}{N^2L}= X\left[X^2-1 + \beta_3^2\left(Y^2-q^2\right) + \frac{H^2}{L^2}\right] \  , \label{eq1}
\end{equation}
\begin{equation}
 \frac{(N^2LY')'}{N^2L}= Y\left[\beta_3^2\left(X^2-1\right) + \beta_2^2\left(Y^2-q^2\right) + \frac{H^2}{L^2}\right], \label{eq2}
\end{equation}
\begin{equation}
 \frac{L}{N^2} \left(\frac{N^2H'}{L}\right)' = \alpha \bigl(X^2 + Y^2)H  \  .  \label{eq3}
\end{equation}
As to the Einstein equations (\ref{eqEinstein}), we obtian:
\begin{equation}
 \frac{\left(LNN'\right)'}{N^2L} = \gamma \left[\frac{H'^2}{2\alpha L^2} -\frac{1}{4}\left(X^2-1\right)^2 -\frac{\beta_2^2}{4}\left(Y^2-q^2\right)^2-
 \frac{\beta_3^2}{2}\left(X^2 -1\right)\left(Y^2-q^2\right)\right]  \label{eq4}
\end{equation}
and
\begin{equation}
\frac{\left(N^2L'\right)'}{N^2L} = -\gamma \left[\frac{H'^2}{2\alpha L^2} + \left(X^2
+Y^2\right)\frac{H^2}{L^2}+\frac{1}{4}\left(X^2-1\right)^2  
 +\frac{\beta_2^2}{4}\left(Y^2-q^2\right)^2 + \frac{\beta_3^2}{2}\left(X^2 -1\right)\left(Y^2-q^2\right) \right]  \  .  \label{eq5}
\end{equation}
The primes in the above equations denote derivatives with respect to $x$.

At this point we would like to notice that the set of differential equation above, Eq.s  \eqref{eq1}-\eqref{eq5}, 
 reduces itself to the corresponding one for the 
Abelian Higgs model \cite{Brihaye-Lubo},  by taking $\beta_2=\beta_3=0$ and setting  one of the 
Higgs field equal to zero. Because one of our objective is to compare our results with the corresponding one for the Abelian 
case,  we shall take, when necessary, the bosonic field $\chi=0$, which, in terms of dimensionless functions, corresponds to take $Y=0$.

\subsection{ Boundary conditions}

The requirements of regularity at the origin and topologically stable solutions,  leads to the following boundary conditions 
for the matter and gauge fields
\begin{equation}
     H(0) = 1; \quad H(\infty) = 0 \  , \label{eqbound1}
\end{equation}
\begin{equation}
     X(0) = 0, \quad X(\infty) = 1,  \quad  Y(0) = 0,  \quad Y(\infty) = \frac{\eta_2}{\eta_1} = q \label{eqbound2}
\end{equation}
and for the metric fields
\begin{equation}
     N(0) = 1, \quad N'(0) = 0 \  , \quad L(0) = 0, \quad L'(0) = 1 \  . \label{eqbound3}
\end{equation}
    
The energy density per unity of lenght is given by: 
\begin{equation}
 \varepsilon = \int \sqrt{^{(3)}g} \ T^{t}_{t}d\rho d\phi \  ,  \label{energy}
\end{equation}
where $^{(3)}g$ is the determinant of the (2+1) dimensional metric given in  \eqref{ds} by taking $dz=0$,  and $T^0_0$ is  the 
00-component of the energy-momentum tensor. Therefore, 

\begin{eqnarray}
\label{energy}
 \varepsilon & = & 2\pi\eta_1^2\int_0^{\infty}NL\Biggl[\frac{H'^2}{2\alpha L^2} + \frac{1}{2}\left(X'^2 + Y'^2\right) \nonumber
 +\frac{1}{2}\left(X^2 + Y^2\right)\frac{H^2}{L^2} + \frac{1}{4}\left(X^2-1\right)^2 \nonumber \\  
& & +\frac{\beta_2^2}{4}\left(Y^2-q^2\right)^2 +\frac{\beta_3^2}{2}\left(X^2-1\right)\left(Y^2-q^2\right)\Biggr]dx \  .\label{eq31}
\end{eqnarray}
The $T^{\rho}_{\rho}$ and $T^{\phi}_{\phi}$ components are, respectively,
\begin{eqnarray}
T^{\rho}_{\rho}  & = & \eta_1^4\lambda_1\Biggl[-\frac{H'^2}{2\alpha L^2} - \frac{1}{2}\left(X'^2 + Y'^2\right) \nonumber   
 +\frac{1}{2}\left(X^2 + Y^2\right)\frac{H^2}{L^2} + \frac{1}{4}\left(X^2-1\right)^2 \nonumber \\      
& &+\frac{\beta_2^2}{4}\left(Y^2-q^2\right)^2 +\frac{\beta_3^2}{2}\left(X^2-1\right)\left(Y^2-q^2\right)\Biggr] \  , \label{eq32}
\end{eqnarray}
\begin{eqnarray}
T^{\phi}_{\phi}  & = & \eta_1^4\lambda_1\Biggl[-\frac{H'^2}{2\alpha L^2} + \frac{1}{2}\left(X'^2 + Y'^2\right)   
 -\frac{1}{2}\left(X^2 + Y^2\right)\frac{H^2}{L^2} + \frac{1}{4}\left(X^2-1\right)^2 \nonumber \\      
& &+\frac{\beta_2^2}{4}\left(Y^2-q^2\right)^2 +\frac{\beta_3^2}{2}\left(X^2-1\right)\left(Y^2-q^2\right)\Biggr]  \label{eq33}
\end{eqnarray}
and $T^z_z = T^t_t$.

\section{ Numerical Results}
\label{Numerical}
In the following we shall analyze our system. To do that we integrate numerically 
the equations (\ref{eq1}) - (\ref{eq5})  with boundary conditions (\ref{eqbound1} - \ref{eqbound3}) 
by using the ODE solver COLSYS \cite{Colsys}. Relative errors of the functions
are typically on the order of $10^{-8}$ to $10^{-10}$ (and sometimes even better).

As we have  said, our objective in this section is to  analyze numerically the behavior of the solutions of the non-Abelian gravitating string
by  specifying  the set of physical parameters of the system; moreover, we are also interested to compare these  behaviors with the corresponding
one for the Abelian gravitating strings, observing, separately, the influence of each system on the geometry of the spacetime. 
 
\begin{figure}[h!]
\begin{center}
\subfigure[]{\label{na_XYH}
\includegraphics[width=8cm,height=7cm, scale=0.8, angle = 0]{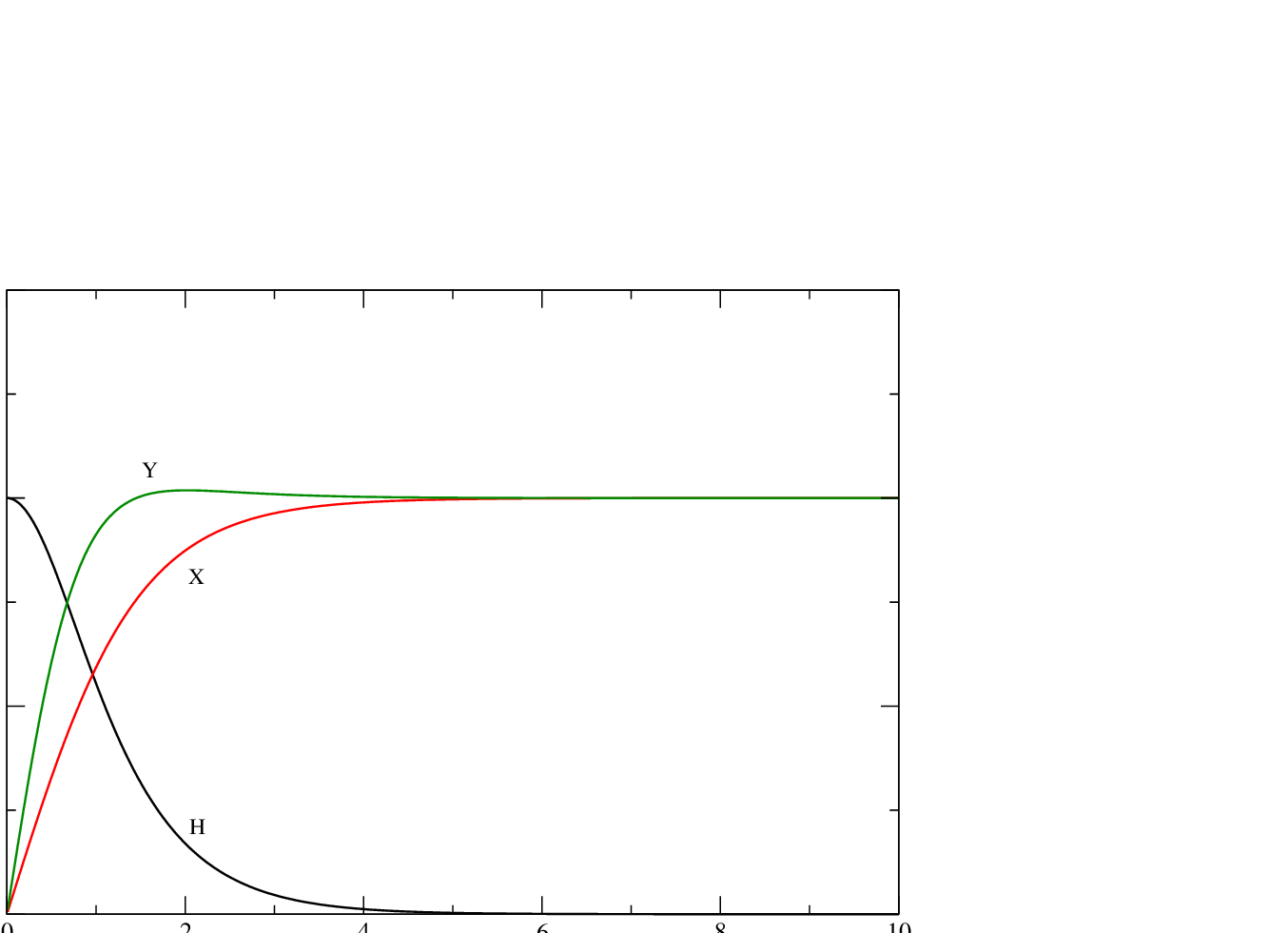}}
\subfigure[]{\label{na_NL}
\hspace{0.5cm}
\includegraphics[width=8cm, height=7cm,scale=0.8,angle = 0]{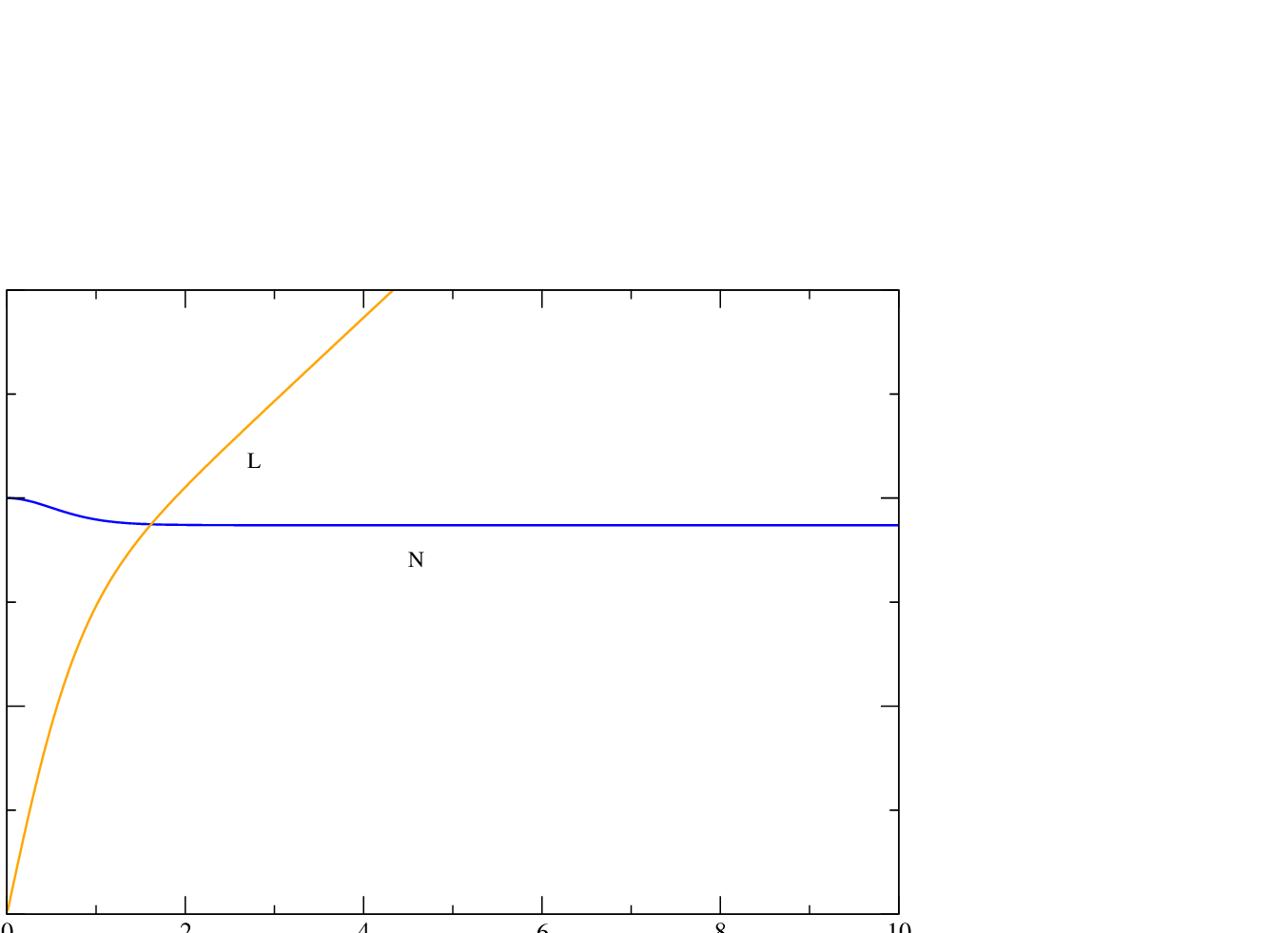}} \\
\end{center}
\caption{\label{non_abelian} { \bf Non-Abelian strings}: In the left panel we present the behavior for the Higgs and gauge fields 
as functions of $x$ considering the parameters  $\alpha=1.0$ $\gamma=0.6$, $\beta_2= 2.0$, $\beta_3=1.0$
and $q=1.0$. In the right panel we present the behavior of the metric functions.}    
\end{figure}

Let us first analyze the behaviors of the Higgs and gauge fields besides the metric functions for the non-Abelian string system. In order to do that 
we construct the plots of these fields as functions of the the dimensionless variable, $x$. Our results are exhibited in Fig. \ref{non_abelian}.
In the left panel we present the behaviors of the Higgs,  $X$ and $Y$, and gauge fields, $H$,  considering $\alpha=1.0$, $\gamma=0.6$,
 $\beta_2=2.0$, $\beta_3=1.0$ and $q=1$. The behaviors of the metric fields, $N$ and $L$,
 are exhibited in the right panel.
 
\begin{figure}[h!]
\begin{center}
\subfigure[]{\label{a_XH}
\includegraphics[width=8cm,height=7cm, angle = 0]{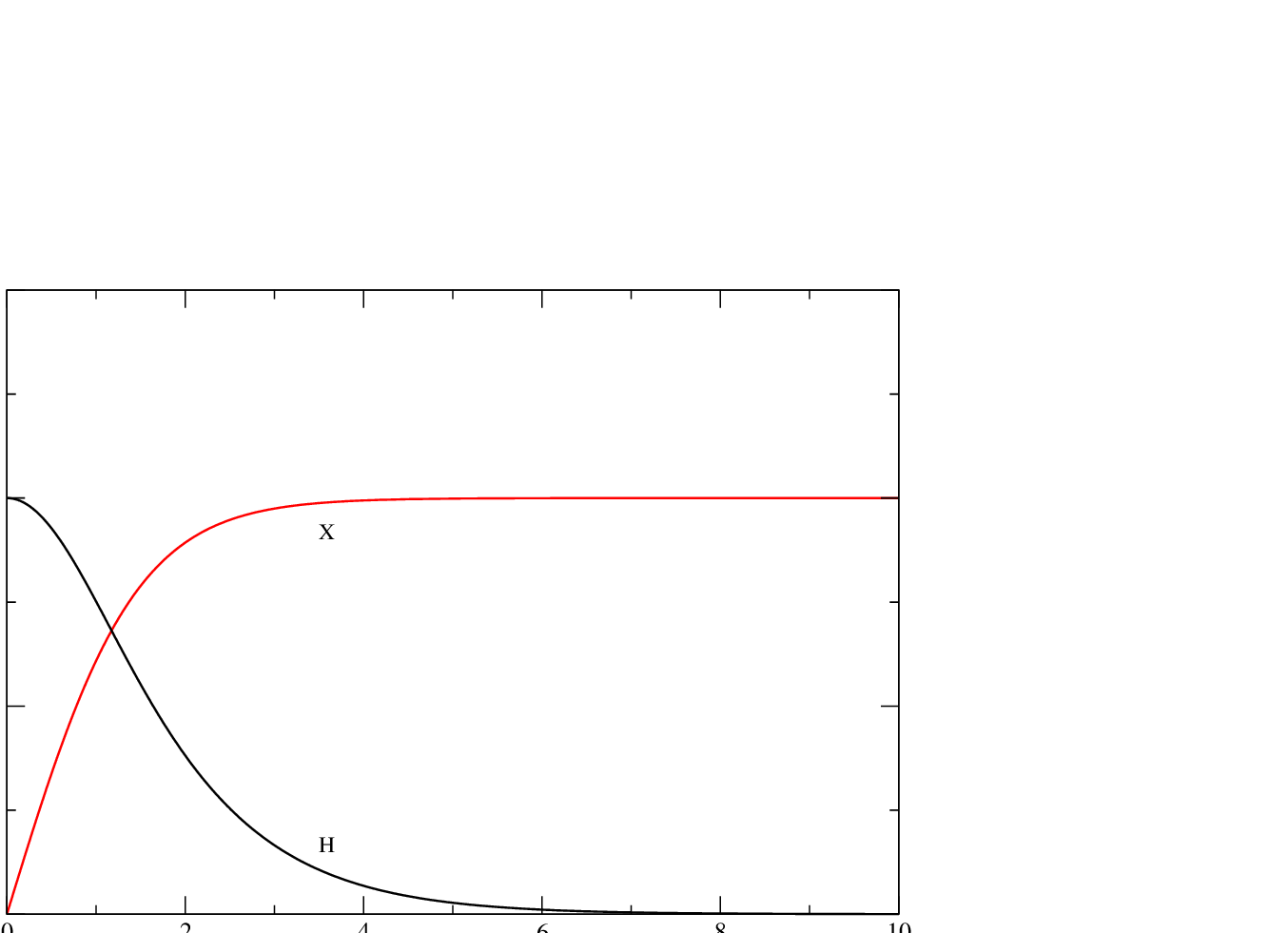}}
\subfigure[]{\label{a_LN}
\hspace{0.5cm}
\includegraphics[width=8cm,height=7cm, angle = 0]{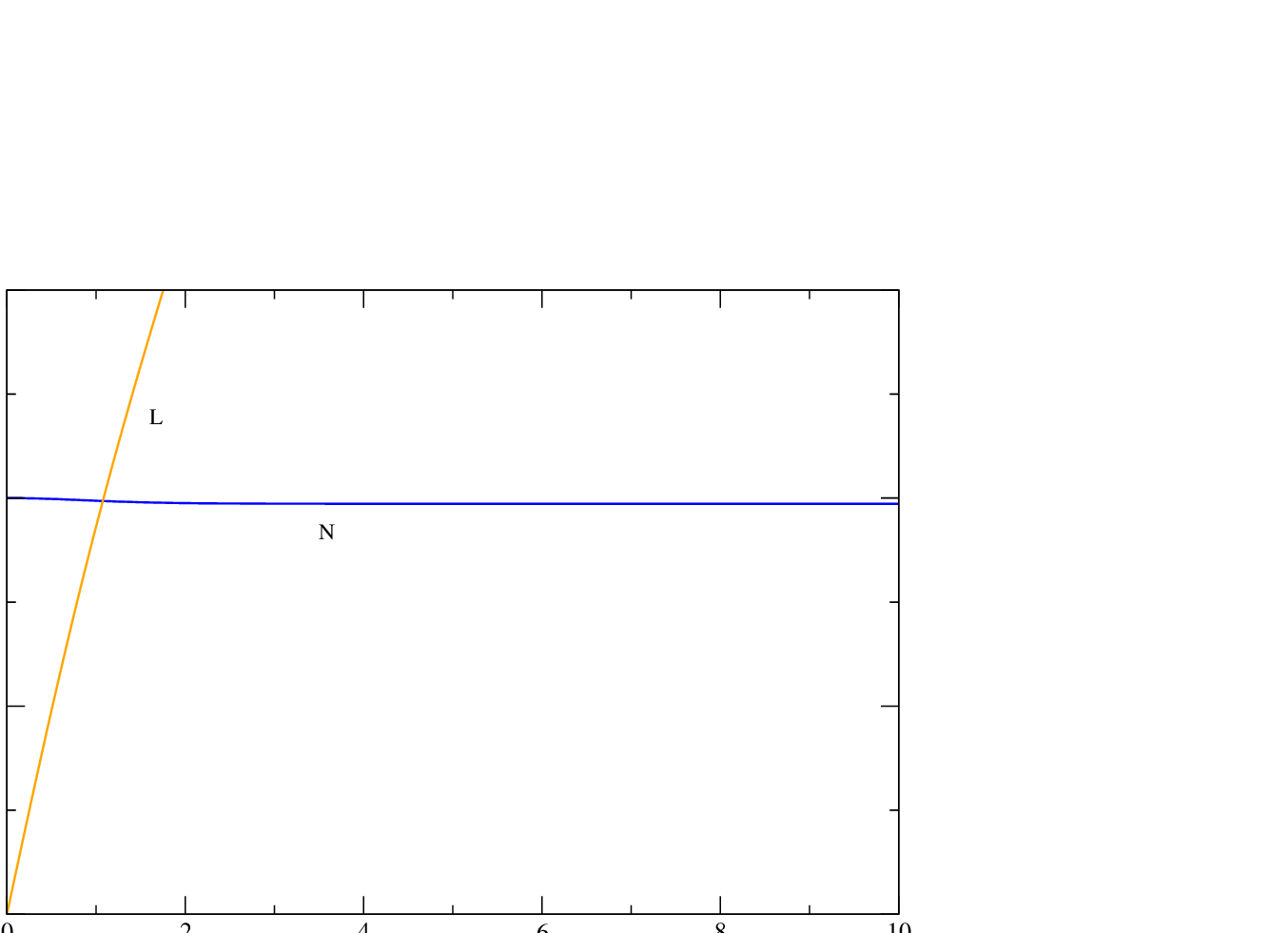}} \\
\end{center}
\caption{\label{abelian} {\bf Abelian strings}: In the left panel we present the behavior for the Higgs and gauge fields as functions of $x$,
considering the parameters  $\alpha=1.0$, $\gamma=0.6$. In the right panel, we present the behavior of the metric functions.}    
\end{figure}

Concerning now the behaviors of the matter, gauge fields and the metric functions for the Abelian string, we adopt the procedure
already explained. So, we plot the fields $X$ and $H$, besides the metric fields, $L$ and $N$,  as functions of $x$.
In the the left panel of Fig. \ref{abelian} we present the behaviors for the Higgs and gauge fields considering $\alpha=1.0$ and $\gamma=0.6$. 
The corresponding  behaviors for the  metric functions are shown in the right panel.
 
\begin{figure}[h!]
\begin{center}
\includegraphics[width=8cm,height=7cm]{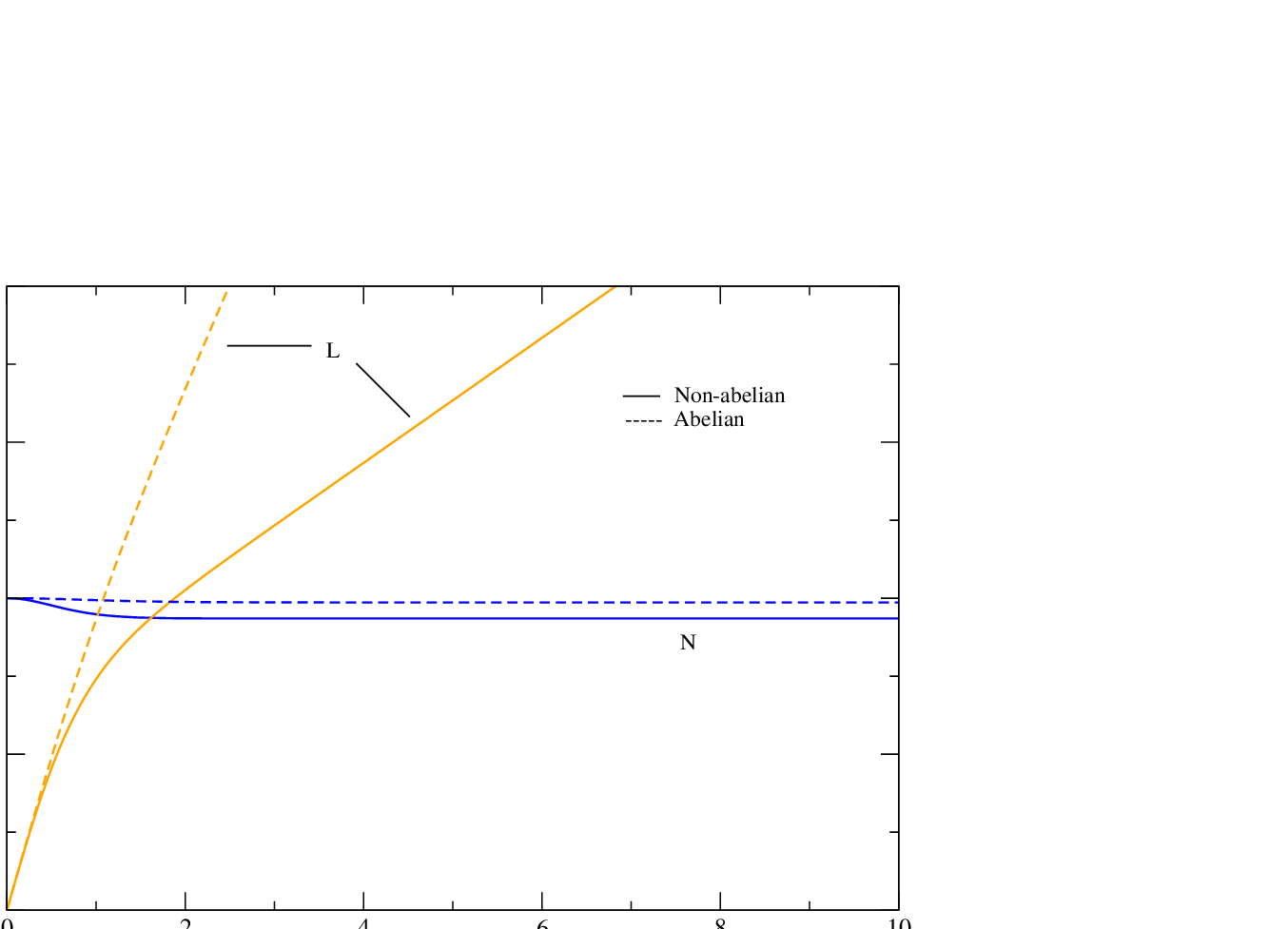}
\end{center}
\caption{This figure shows the comparison between the behavior of the metric functions as functions of $x$, 
considering $\alpha=1.0$, and $\gamma=0.6$. For the non-Abelian case, we have taken  $\beta_2=2.0$, $\beta_3=1.0$ and $q=1.0$.\label{A-NA}}
\end{figure}

The comparison of the behaviors for the metric functions, $L$ and $N$, as functions of $x$ for both,  
non-Abelian and Abelian strings,  are exhibited in Fig. \ref{A-NA},
considering  $\alpha=1.0$, $\gamma=0.6$, $\beta_2=2.0$, $\beta_3=1.0$ and $q=1.0$ for the non-Abelian string
 and $\alpha=1.0$, $\gamma=0.6$ for the Abelian string.

Both, non-Abelian and Abelian string solutions, provide planar angle deficit in the corresponding spacetime. This quantity is given by analyzing the 
slop of $L$ for points very far from the string's core. The planar angle deficit is given by:
\begin{equation}
 \label{deficit}
 \delta=2\pi(1-L'(\infty))  \  . 
\end{equation}
Moreover, the linear energy density is obtained by integrating  \eqref{eq31}. Taking for
the parameters the values used  in the previous plots,
we have found for the planar angle deficits and energy densities per unity of $2\pi\eta_1^2$,
 respectively, the following results:
\begin{equation}
\delta_{NA}\approx 0.7998 \  {\rm and} \  \delta_A\approx 0.3493 \  
\end{equation}
and\footnote{The sub-scripts $NA$ and $A$ refer to the non-Abelian and Abelian strings, respectively.}
\begin{equation}
\varepsilon_{NA}/(2\pi\eta_1^2)\approx 1.2856 \  {\rm and} \ \varepsilon_{A}/(2\pi\eta_1^2)\approx1.1646 \  . 
\end{equation}

Another relevant analysis associated with the non-Abelian strings concerns the behavior of their energy densities. Two
independent analysis have developed  by us: The energy density per units of $2\pi\eta_1^2$ as function of $\beta_3$ and $\gamma$. In the left
panel of the Fig. \ref{energy}, we exhibit this behavior as function of $\beta_3$ considering $\gamma=0.6$. In the
right panel we present this behavior as function of $\gamma$ considering $\beta_3=1.0$. For both plots we have
taken $\alpha=1.0$, $\beta_2=2.0$ and $q=1$.
\begin{figure}[!h]
\begin{center}
\subfigure[]{\label{fig4a}
\includegraphics[width=8.5cm,height=8cm, angle =0]{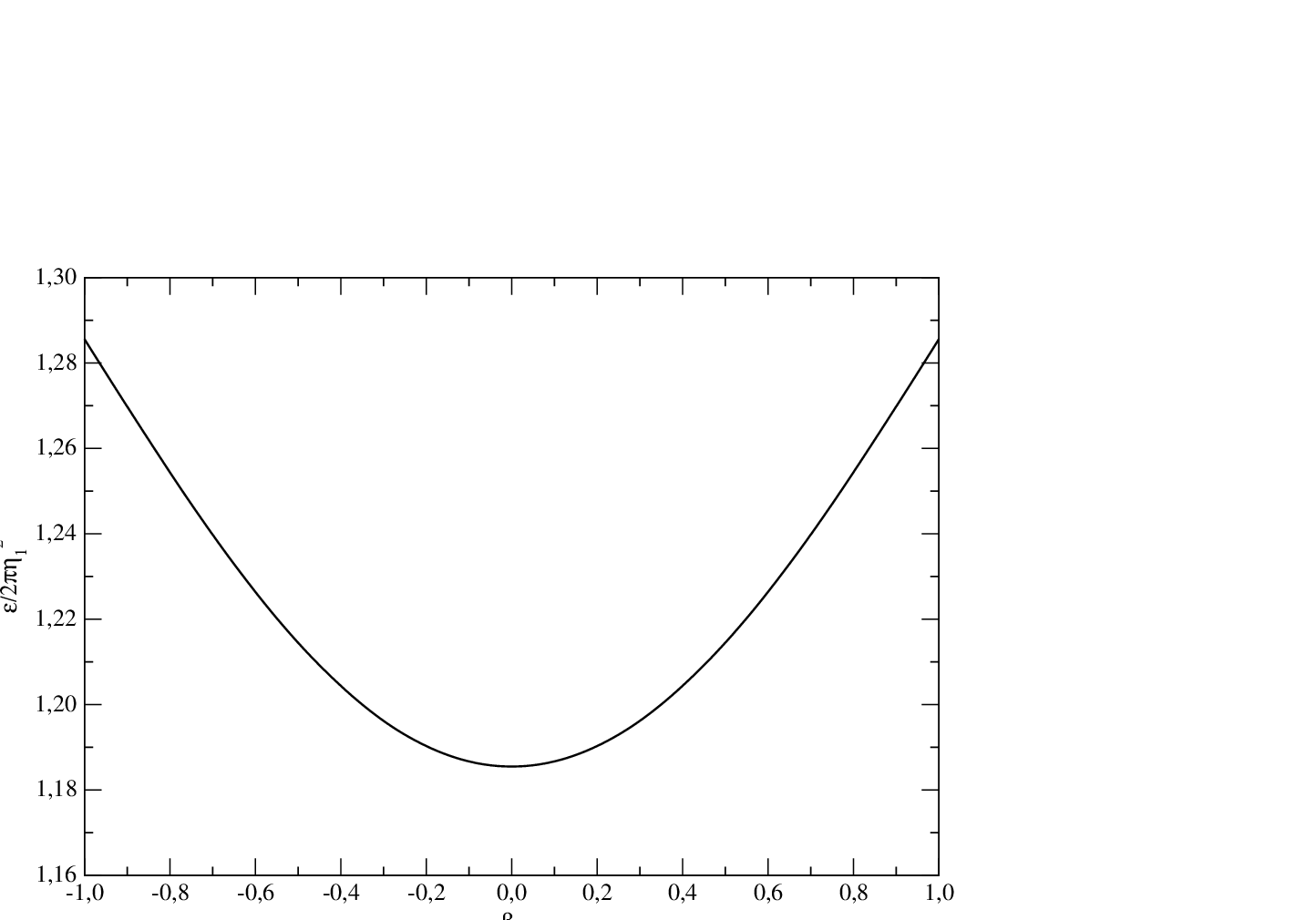}}
\subfigure[]{\label{fig4b}
\hspace{0.5cm}
\includegraphics[width=8.5cm,height=8cm, angle =0]{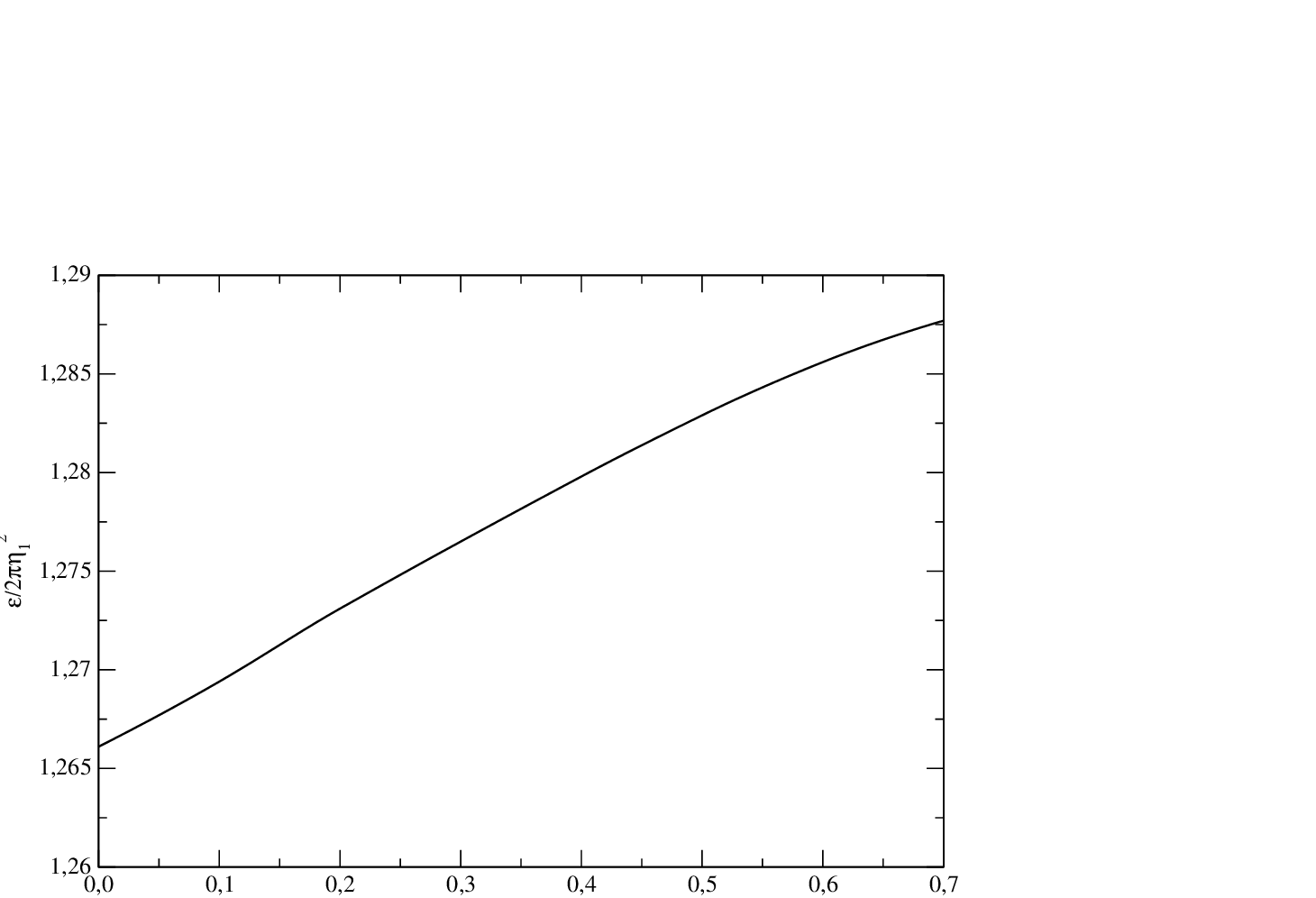}} \\
\end{center}
\caption{\label{energy} In the left panel we present the behavior of the energy density per unit of
$2\pi\eta_1^2$ as function of $\beta_3$ considering $\gamma=0.6$. In the right panel we present the behavior of the energy density
per unit of $2\pi\eta_1^2$  as function of $\gamma$, considering $\beta_3=1.0$. In both graphs we have taken  
$\alpha=1.0$, $\beta_2=2.0$ and $q=1$.}    
\end{figure}

\begin{figure}[h!]
\begin{center}
\subfigure[]{\label{fig5a}
\includegraphics[width=8.5cm,height=8cm, angle =0]{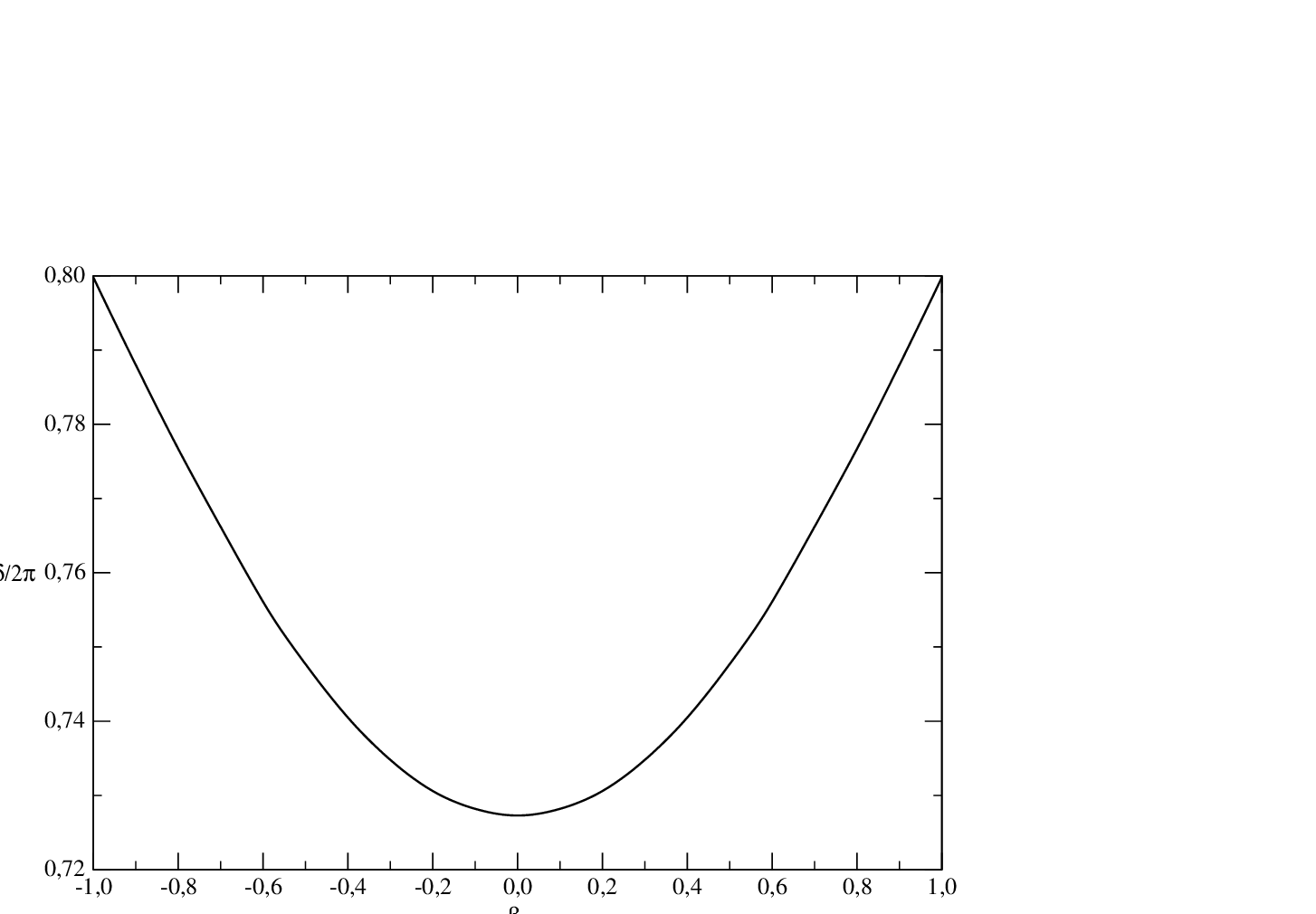}}
\subfigure[]{\label{fig5b}
\hspace{0.5cm}
\includegraphics[width=8.5cm,height=8cm, angle =0]{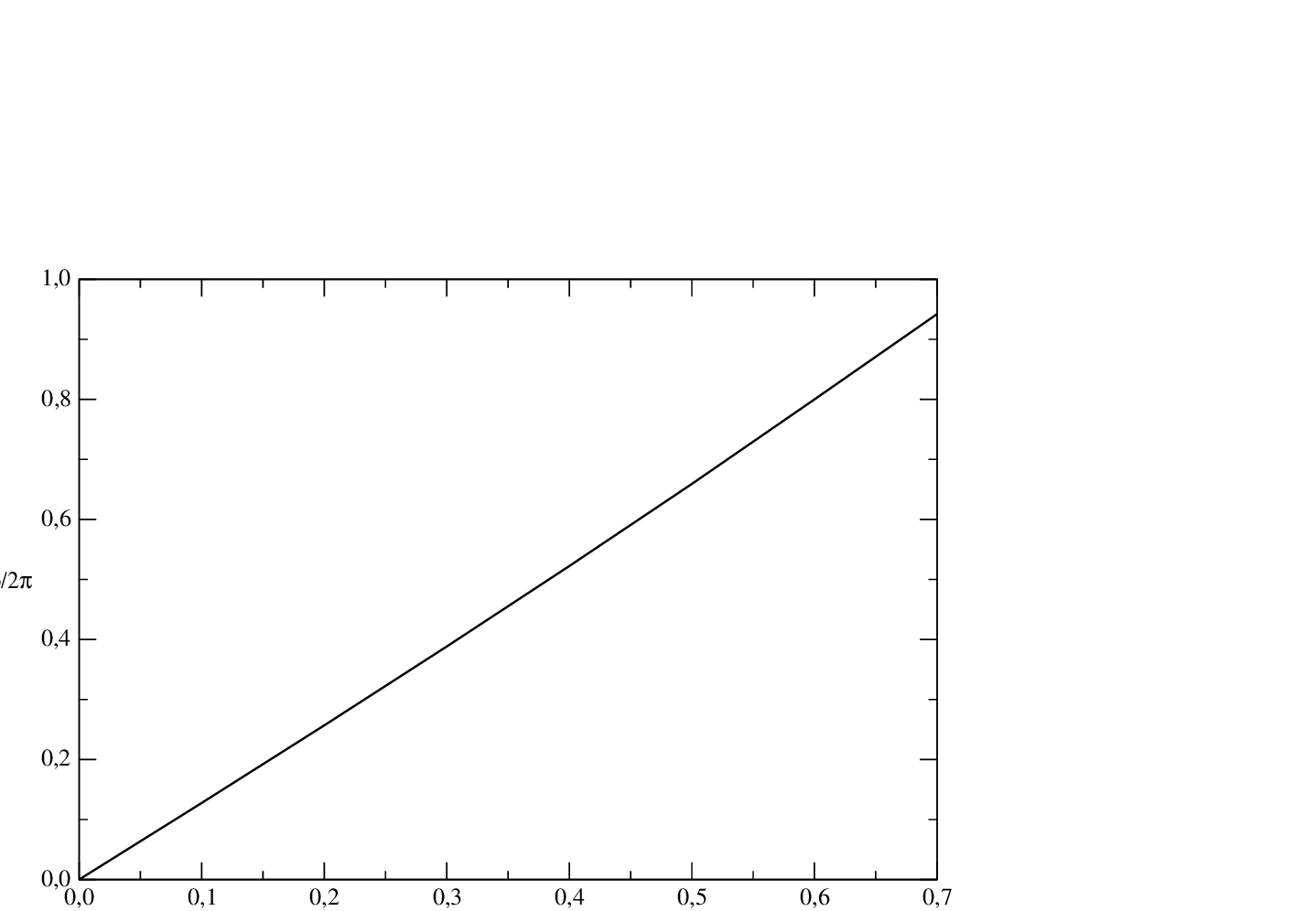}} \\
\end{center}
\caption{\label{deficit-1} In the left plot we present the behavior of the the planar angle deficit in units of $2\pi$
as function of $\beta_3$ considering $\gamma=0.6$. In the right plot we present this behavior
as function of $\gamma$, considering $\beta_3=1.0$. In both plots we have taken  
$\alpha=1.0$, $\beta_2=2.0$ and $q=1$.}    
\end{figure}

The behaviors of the planar angle deficit in units of $2\pi$, $\delta/2\pi$, as function of $\beta_3$ and $\gamma$, are presented in 
Fig. \ref{deficit-1}. In the left plot we present  $\delta/2\pi$ as function of $\beta_3$ considering $\gamma=0.6$. In the
right plot we present this behavior as function of $\gamma$ considering $\beta_3=1.0$. For both plots we have
taken $\alpha=1.0$, $\beta_2=2.0$ and $q=1$.

\begin{figure}[h!]
\begin{center}
\includegraphics[width=10cm,height=8cm]{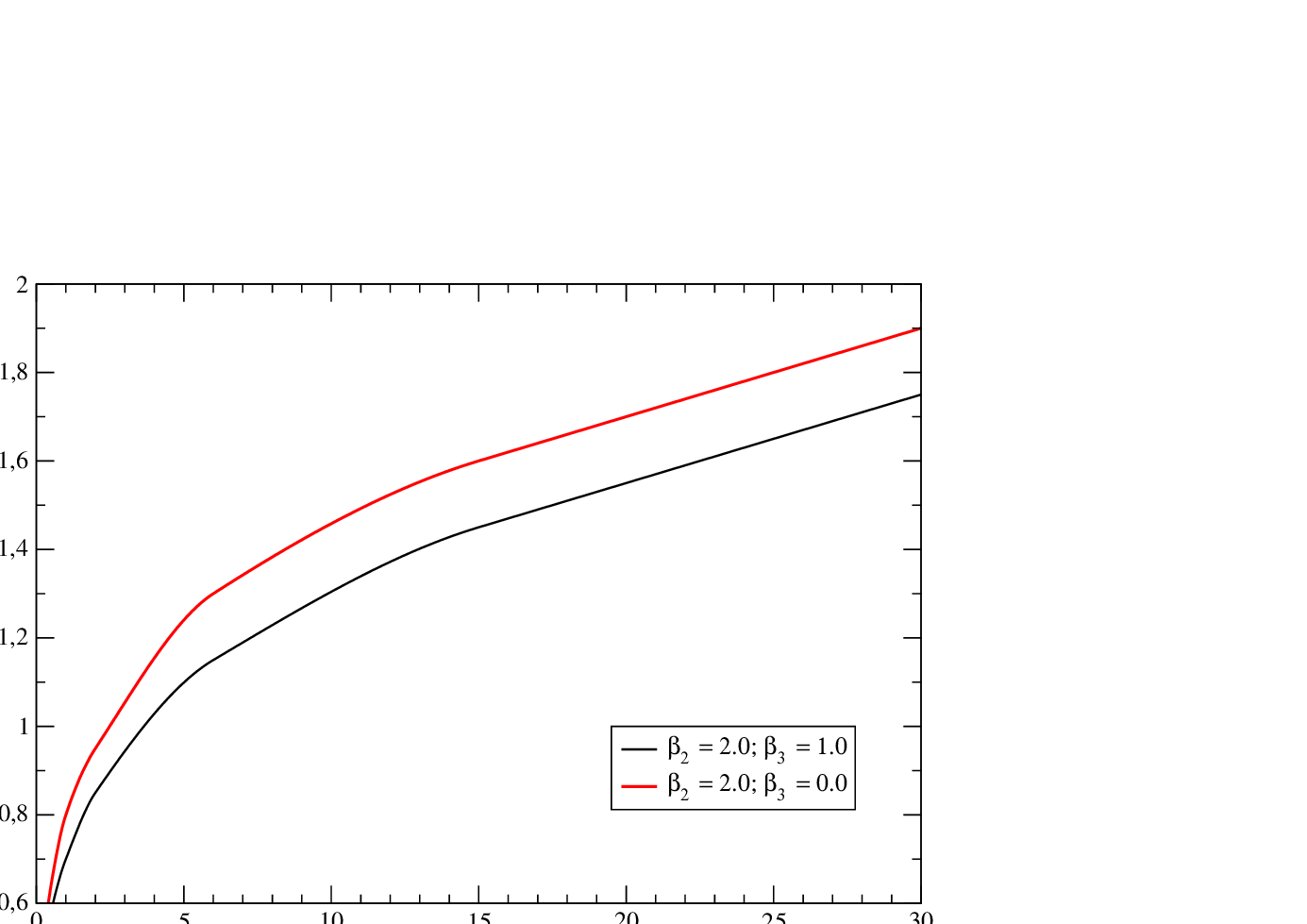}
\end{center}
\caption{This  figure shows the the regions in the  $(\alpha-\gamma)$ parameter space, which contains regular 
non-Abelian string solutions. Our plots were  developed considering two distinct values of $\beta_3$ (see the graph). 
For both cases, we consider $q=1.0$ and $\beta_2=2.0$. \label{phase}}
\end{figure}

The solutions that we are analyzing in this paper, named regular strings, are the ones that 
present the planar angle deficit, $\delta$,  smaller than $2\pi$. The planar angle deficit is a measurement of the 
the modification on the geometry of the spacetime caused by gravitational  interaction of the system. 
To obtain regular strings is required an optimal choice of the parameters $\alpha$, $\gamma$, 
$\beta_2$, $\beta_3$ and $q$. For fixed values of $\beta_2$, $\beta_3$ and $q$, it is possible 
to determine regular solutions examining the region below the curve in the $(\alpha-\gamma)$ parameter space. 
However, from equation \eqref{eq31}, the energy density decreases as $\alpha$ is increased. 
Moreover, as shown in the Fig. \ref{fig4b}, the energy density 
increases with  $\gamma$. Therefore, regular strings are obtained until the critical 
value of $\gamma$ is achieved, $\gamma_{cr}$. 
This $\gamma_{cr}$ is described by the curve in the $(\alpha-\gamma)$ parameter space. 
If $\beta_3=0.0$, $\gamma_{cr}$ is greater than in the 
case of $\beta_3\neq0.0$ because there is no contribution to energy density from the coupled bosonic sectors. This fact is showed
in Fig. \ref{phase}, where we have considered  $\beta_2=2.0$ and
$q=1$, for $\beta_3=0.0$ and $1.0$.

\section{Conclusion}
\label{Concl}

In this paper we have investigated gravitating non-Abelian, $SU(2)$,  Higgs model of cosmic strings. In order to have non-Abelian
stable topological string it is necessary the presence of of two bosonic iso-vectors. As to the interaction potential,
besides contains forth power terms in the fields, 
it may present a direct coupling interaction between the bosonic sectors. The numerical analysis were developed 
considering specific values of the physical parameters of the model. 

One of our objective was to compare the behaviors of the metric functions corresponding to the non-Abelian and Abelian 
gravitating strings as function of the dimensionless variable, $x$. These behaviors were exhibited, separately,  
in Fig. \ref{non_abelian} and  Fig.  \ref{abelian}.  In Fig \ref{A-NA}, we have displayed the behavior of the metric functions, $N$ and $L$,
for both strings. We can notice that, for specific values of the parameter\footnote{We have considered 
 $\alpha=1.0$ and $\gamma=0.6$ for both cases, and $\beta_2=2\beta_3=2.0$
and $q=1$ for the non-Abelian string.},  the non-Abelian string  provides a larger planar angle deficit than the corresponding Abelian one, 
although the corresponding energy density is larger. In fact, by our results,  the planar angle deficit is approximately $2.290$ bigger and
the corresponding energy density is also bigger,  approximately $1.104$.  

As to the non-Abelian string, we have analyzed the behavior of the energy density as function of $\beta_3$ and $\gamma$.  In 
figure \ref{fig4a} we can observe that its minimum value occurs for $\beta_3=0$, and in the  figure \ref{fig4b}
we see that the energy density  increases with $\gamma$. 

The analysis of the planar angle deficit as function of $\beta_3$ and $\gamma$ are exhibited in Fig. \ref{deficit-1}. In the figure \ref{fig5a}
we can see that the minimum value for this quantity happens for $\beta_3=0$; of course, in this analysis, it is necessary that $\gamma>0$.

We have also provided in Fig. \ref{phase}, a graph that exhibit a region in the $(\alpha-\gamma)$ parameter space where regular
strings can be formed for $\beta_2=2.0$ and $q=1$ for two specific values of $\beta_3$. This region is below the curved line in the
plot. As we can see, considering $\beta_3=0$, for a given $\alpha$ the critical value for $\gamma$ which allows regular
string is bigger than for $\beta_3=1$. This fact is consequence of the dependence of the energy density, Eq. \eqref{eq31},  with $\beta_3$.
Because the planar angle deficit increases with $\varepsilon$, the planar angle deficit depends on $\beta_3$; so
vanishing this parameter, to attain $\delta/2\pi$ near unity is required larger value of $\gamma$.

In great part of the  analysis developed in this paper we have considered parameters that provide $\Delta>0$; however, in the  Appendix,  we analyze 
a special case of non-Abelian string where this condition is not fulfilled. This solution presents some similarity with the BPS solution of the
Abelian Higgs string, i.e., it presents the metric function $N$ equal to unity, and allows to reduce the set of differential equation to a simpler one.
We also analyzed in the Appendix the behavior of the BPS Abelian-Higgs string.
\\

{\bf Acknowledgment:}  E.R.B.M. would like to acknowledge CNPq for partial financial support. A. de P. S. would like to acknowledge the Universidade
Federal Rural de Pernambuco. Also the authors would like to acknowledge Betti Hartmann for her valuable contributions during all steps of this paper.

\appendix
\section{The Special Case}

As it is well known, the Abelian-Higgs system presents the BPS solution for the fields equation. This solution is characterized by the fact of  be solution
of a set of first  order differential equation  which minimize the  energy density of the string. This configuration is achieved by taking
$\alpha=2.0$ in the equations of motion and assuming $N=1$ \cite{Brihaye-Lubo}.
The BPS limit for the Abelian Higgs model can be obtained from our system of differential  equations, 
by taken $\beta_2=\beta_3=0$, $\alpha=2.0$ and assuming the field $Y=0$ and the metric function $N=1$
equal to the unity. In this case the set of differential equations reduce to:
\begin{equation}
 H' = L\left(X^2  -1\right) \  ,   L'' = -\gamma L \left[\frac{ X^2 H^2}{L^2} +\frac{\left(X^2  -1\right)^2}2\right ] 
\end{equation}
and
\begin{equation}
  X' = X\frac{H}{L}  \  .
\end{equation}

Here, for the non-Abelian string case, 
we wish to analyze a special solution of the fields equation that present characteristic similar to the Abelian  BPS one; however, for this case,  
we have $\Delta=\lambda_1\lambda_2-\lambda_3^2=0$.  In order to construct this solution we shall substitute $\alpha = 2.0$, $\beta_i = 1.0$, 
$N=1$ and $q = 1.0$ into the equations (\ref{eq4}) and (\ref{eq5}). So,  we get:

\begin{equation}
 H' = L\left(X^2 + Y^2 -2\right) \label{A3}
\end{equation}
and
\begin{equation}
 L'' = -\gamma L \left[\left(X^2 + Y^2\right)\frac{H^2}{L^2} +\frac{1}{2}\left(X^2 + Y^2 -2\right)^2\right ] \  . \label{A4}
\end{equation}
Moreover, taking $\alpha = 2.0$ into Eq. (\ref{eq3}), two particular independent linear differential
equations can be obtained. They are:
\begin{equation}
 X' = X\frac{H}{L} \ {\rm and} \  Y' = Y\frac{H}{L} \  . \label{A5}
\end{equation}
With these conditions, we can verify that the radial and azimuthal components of the energy-momentum
tensor vanish:

\begin{equation}
 T^{\rho}_{\rho} = T^{\phi}_{\phi} = 0 \ .
\end{equation}
As to the energy density and axial tension, by using equations \eqref{A3} to \eqref{A5}, they can be expressed by:
\begin{equation}
 T^z_z = T^t_t = \frac{\eta_1^4\lambda_1}{2L}\left[H( X'^2 + Y'^2-2)\right]' \  ,
\end{equation}
where the derivative is with respect to $x$.

\begin{figure}[h!]
\begin{center}
\subfigure[][]{\label{na_BPS}
\includegraphics[width=8cm, angle = 0]{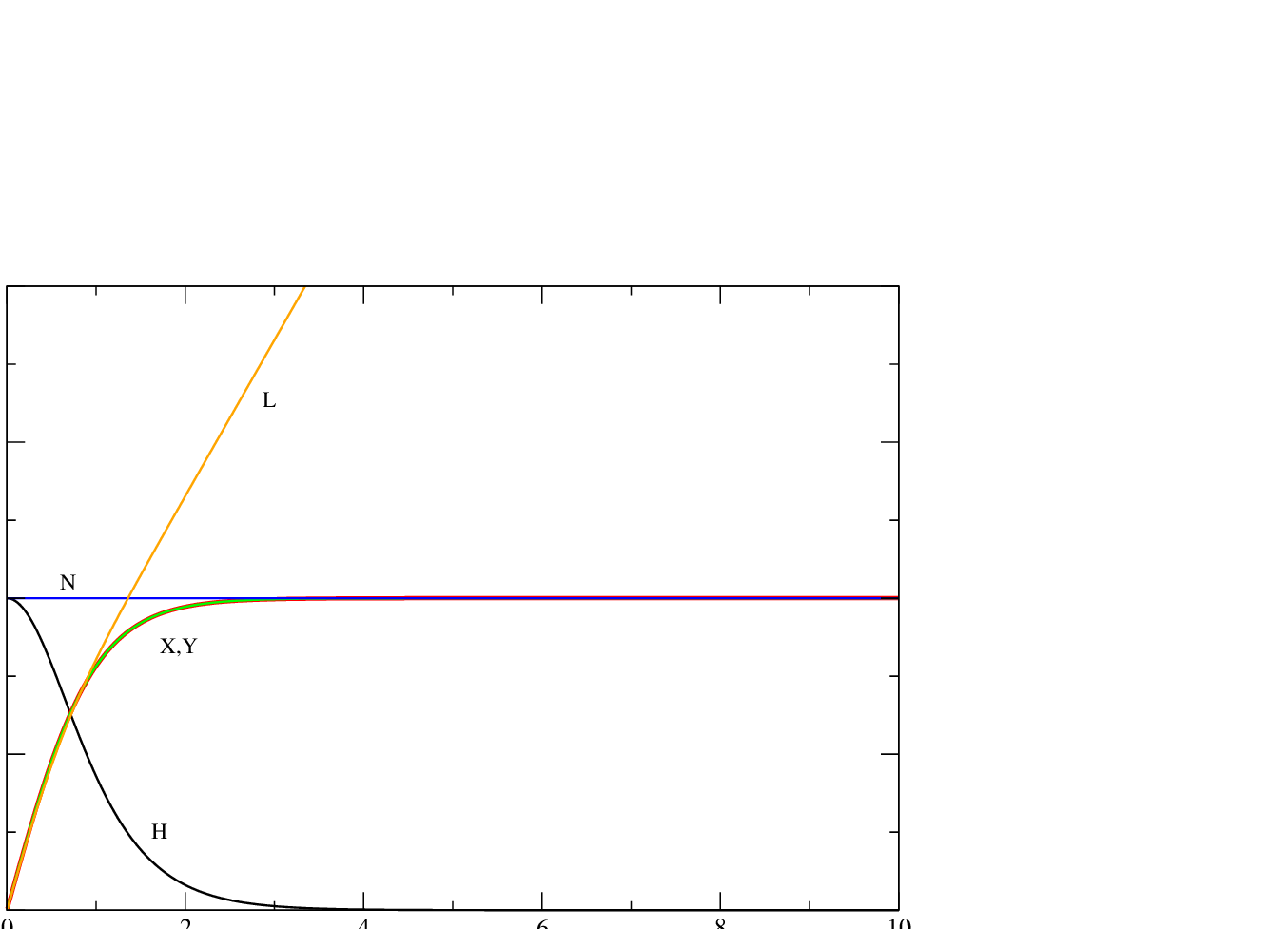}}
\subfigure[][]{\label{a_BPS}
\hspace{0.5cm}
\includegraphics[width=8cm, angle = 0]{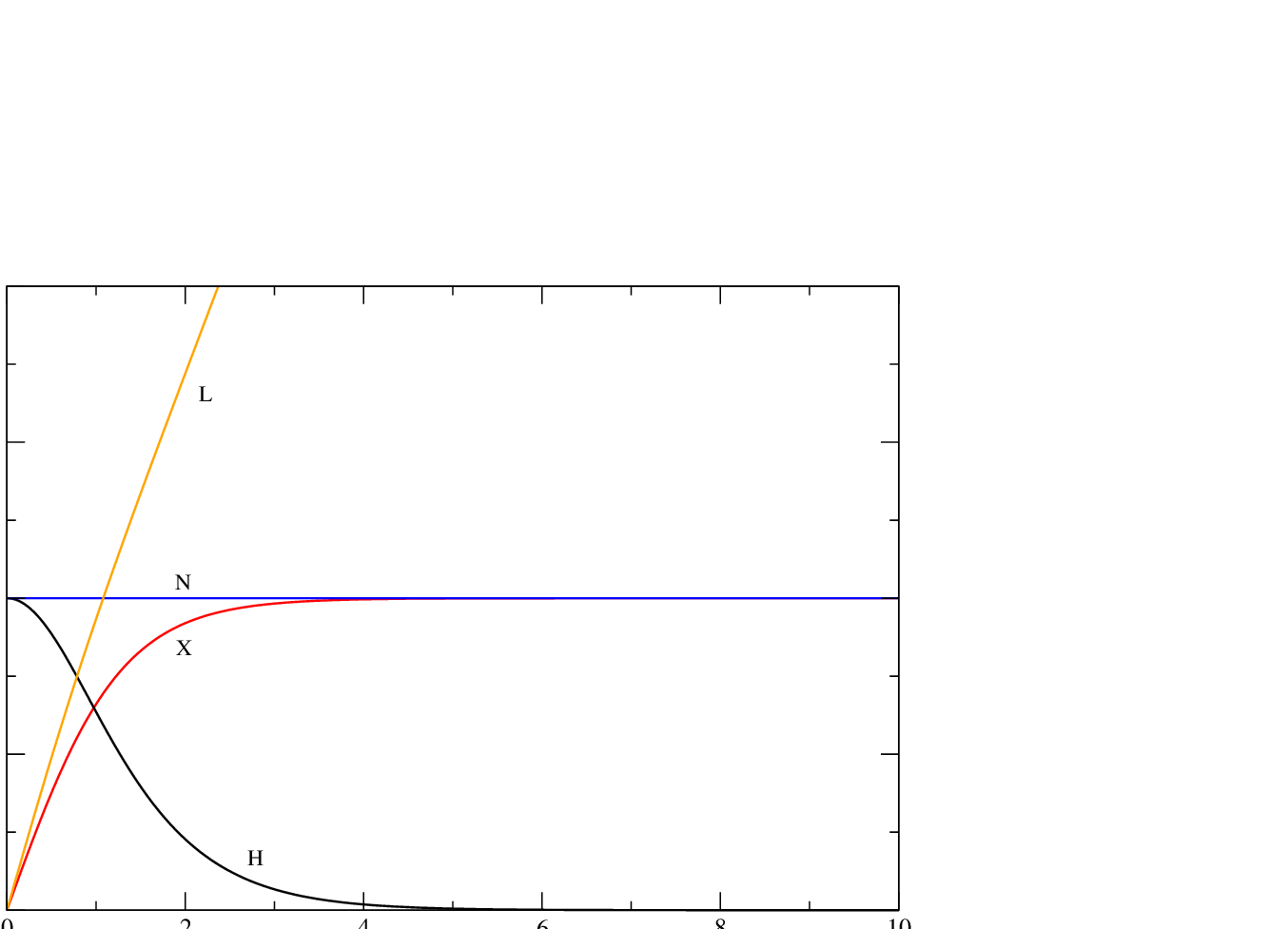}} \\
\end{center}
\caption{\label{Non_Abelian} In the left panel we present the behaviors for the Higgs and gauge fields, 
and for the metric functions considering $\gamma=0.6$, $\alpha=2.0$, $\beta_2=\beta_3=1$
and $q=1$, for the non-Abelian special case. In the right panel, we present the behaviors of the Higgs and gauge 
fields, and the metric functions for the Abelian BPS string, considering $\gamma=0.6$ and $\alpha=2.0$.}    
\end{figure}

Our numerical analysis for both, non-Abelian and Abelian, solutions above mentioned are displayed in  Fig. \ref{Non_Abelian}.
In the left panel, we present the behaviors of the Higgs and gauge fields, and the metric functions, for this special case for the non-Abelian string,
considering $\alpha=2.0$, $\beta_2=\beta_3=1$ and $q=1$. In the right panel we exhibit the behaviors for the Higgs and gauge fields, and also for
the metric functions for the  BPS solution of the Abelian strings considering $\alpha=2.0$. 
For both cases we took  $\gamma=0.6$. 

Another point that we want to mention is with respect to the energy per unity length.
For both configurations, the energy density per $2\pi\eta_1^2$ are equal to unity. As to the planar angle 
deficit it has been shown in \cite{Verbin} that for the Abelian BPS solution,
\begin{equation}
\frac\delta{2\pi}=\frac\gamma 2 \   .\label{planar1}
\end{equation}
For the special case of non-Abelian string, we can show by combining \eqref{A3} to \eqref{A5} that
\begin{equation}
L''=-\frac\gamma 2\left[H(X^2+Y^2-2)\right]' \  .
\end{equation}
Integrating the above expression we have,
\begin{equation}
L'(\infty)=1-\gamma \  . 
\end{equation}
Consequently by \eqref{deficit}, we obtain,
\begin{equation}
\frac\delta{2\pi}=\gamma  \   .\label{planar2}
\end{equation}
As we can see, although both solutions presents the same energy density,
 the planar angle deficit associated with the non-Abelian  ''BPS''  configuration is twice bigger  than the deficit for the Abelian BPS string. 
The reason for this fact is because in the non-Abelian  ''BPS''  configuration there are the contribution of
two bosonic sectors against just one bosonic sector for the Abelian BPS solution in the definition of the planar angle deficit. 
Finally we want to mention that our numerical results have exhibited that:
\begin{itemize}
\item Both energies density per $2\pi\eta_1^2$ are equal to unity.
\item Both planar angle deficits are in good agreement with the expressions \eqref{planar1} and \eqref{planar2}. 
\end{itemize}

\end{document}